\RequirePackage{ifpdf}
\documentclass[letterpaper]{JHEP3}
\usepackage{amssymb,amsfonts}
\usepackage{cite}
\usepackage{epsfig}

\newcommand{\R}{{\bf R}}

\newcommand{\CL}{{\cal L}}

\newcommand{\CV}{{\cal V}}

\newcommand{\bv}{{\bf v}}

\newcommand{\bx}{{\bf x}}

\newcommand{\p}{\partial}

\renewcommand{\tilde}[1]{\widetilde{#1}}
\newcommand{\be}{\begin{equation}}
\newcommand{\ee}{\end{equation}}
\newcommand{\bea}{\begin{eqnarray}}
\newcommand{\eea}{\end{eqnarray}}

\newcommand{\muir}{{\mu^{\ }_{\rm IR}}}
\usepackage{graphicx}
\usepackage{latexsym}

\title{Surprises with Nonrelativistic Naturalness}
\author{Petr Ho\v{r}ava
\\ 
Berkeley Center for Theoretical Physics and Department of Physics\\
University of California, Berkeley, CA 94720-7300, USA\\
and\\
Theoretical Physics Group, Lawrence Berkeley National Laboratory\\
Berkeley, CA 94720-8162, USA}
\abstract{We explore the landscape of technical naturalness for nonrelativistic systems, finding surprises which challenge and enrich our relativistic intuition already in the simplest case of a single scalar field.  While the immediate applications are expected in condensed matter and perhaps in cosmology, the study is motivated by the leading puzzles of fundamental physics involving gravity:  The cosmological constant problem and the Higgs mass hierarchy problem.\bigskip\\ 
\noindent
This brief review is based on talks and lectures given at the 2nd LeCosPA Symposium on {\it Everything About Gravity} at National Taiwan University, Taipei, Taiwan (December 2015), to appear in the Proceedings; at the {\it International Conference on Gravitation and Cosmology}, KITPC, Chinese Academy of Sciences, Beijing, China (May 2015); at the Symposium {\it Celebrating 100 Years of General Relativity}, Guanajuato, Mexico (November 2015); and at the {\it 54.\ Internationale Universit\"atswochen f\"ur Theoretische Physik}, Schladming, Austria (February 2016).\bigskip\\
\begin{center}
Published in {\it Int.\ J. Mod.\ Phys.\/} {\bf D}
\end{center}
}
\begin{document}
\baselineskip=13pt
\section{Puzzles of Naturalness}

Some of the most fascinating open problems in the fundamental physics of the Universe can be phrased as puzzles of technical naturalness:  
\begin{itemize}
\item{\bf The cosmological constant problem.} Why is the cosmological constant so small compared to the Planck scale $M_P$ of gravity?  
\item{\bf The Higgs mass hierarchy problem.} Why is the Higgs mass so small compared to the anticipated high scale $M$ in particle physics?  (Here $M$ could be a GUT scale, or some other natural high scale of beyond-standard-model physics -- but it is logical to take $M\sim M_P$, since $M_P$ is the only such high scale that we know to exist for sure). 
\end{itemize}
In cosmology, if one subscribes to slow-roll inflation, another puzzle of naturalness emerges:  Why is the $\eta$ parameter of the inflaton potential so small?  The requirement of technical naturalness \cite{th} follows from some of our most elementary and robust principles of quantum mechanics, quantum field theory and many-body physics, such as causality and spacetime locality.  If violated in Nature, it would suggest disconcerting limitations to the basic principles which have so far been enormously successful across a remarkable range of scales of physical phenomena, from microphysics to condensed matter.  

Notably, all the naturalness puzzles mentioned above involve gravity; it is also intriguing that they can all be phrased as questions about the natural size of various terms appearing in the potential of scalar fields.  It is possible that it is quantum gravity which behaves in ways different from conventional many-body physics, and is ultimately responsible for the apparent violations of technical naturalness.  Before accepting or rejecting this possibility, however, it is worthwhile to adopt technical naturalness as our guiding principle, to look for new symmetries and mechanisms that respect this principle, and to explore and map out the ``landscape of naturalness.'' This is the strategy adopted in this brief review, and in our papers \cite{msb,pol,cmu,nrr} whose main results this paper is based on.    

Puzzles of technical naturalness are not confined only to particle physics, gravity and cosmology.  A beautiful example exists in the field of high-$T_c$ superconductivity:  
\begin{itemize}
\item {\bf $T$-linear resistivity of strange metals.}  In the ``strange metal'' phase above the critical temperature, the resistivity behaves linearly with temperature, across a wide range of scales (at least by condensed matter standards).
\end{itemize}
As was explained beautifully by Polchinski in \cite{joef}, no known interaction could explain this linear dependence in a technically natural way.%
\footnote{At least at the time; since then, the situation has improved somewhat, specifically with the scenarios involving quantum critical points, and particularly those amenable to holographic dual descriptions.  However, the puzzle still persists and awaits its conclusive resolution.}  

\subsection{Technical Naturalness}

Following a period of intuitive developments, the concept of Technical Naturalness was clearly formulated by G. 't Hooft in 1979 in \cite{th}.  The philosophical basis for this principle is clear from the very first sentence of \cite{th}:

\begin{quote}
{\it The concept of causality requires that macroscopic phenomena follow from microscopic equations.}
\end{quote}

Then, 't Hooft further formalizes this philosophy as follows:

\begin{quote}
{\it We now conjecture that the following dogma should be followed:\hfill\break
-- at any energy scale $\mu$, a physical parameter or set of physical parameters $\alpha_i(\mu)$ is allowed to be very small only if the replacement $\alpha_i(\mu)=0$ would increase the symmetry of the system. --}
\end{quote}

A classic example, which will be quite useful in the following, is the first one discussed in \cite{th}:  The relativistic $\lambda\phi^4$ scalar field theory in $3+1$ dimensions, with action  
\be
S=\int d^4x\left(\frac{1}{2}\p_\mu\phi\p^\mu\phi-\frac{1}{2}m^2\phi^2-\frac{\lambda}{4!}\phi^4\right).
\ee
Can the mass $m$ be naturally small?  Yes, but only if $\lambda$ is also correspondingly small: 
\be
\label{epssc}
m^2\sim\varepsilon\mu^2,\qquad \lambda\sim\varepsilon,
\ee
with $\varepsilon\ll 1$ and with $\mu$ a natural scale which makes the dimensions correct.  In this example, the symmetry protecting the mutual smallness of $m^2$ and $\lambda$ is the celebrated ``constant shift'' symmetry, generated by
\be
\delta \phi(x^\mu)=a,
\ee
with $a$ a real constant, and broken by a small amount at the high scale $\mu$.  This ``naturalness scale'' $\mu$ can also be easily extracted from (\ref{epssc}):
\be
\mu\sim\frac{m}{\sqrt{\lambda}}.
\ee
Thus, if we wish for $\mu$ to be much larger than the physical particle mass $m$, the theory must be kept at very weak coupling.  Or, put differently, for a fixed $m$ the theory is technically natural up to the scale $\mu$, but not much higher.

Our third and final quote from \cite{th} sounds both quite visionary and rather ominous: 

\begin{quote}
{\it Pursuing naturalness beyond 1000 GeV will require theories that are immensely complex compared with some grand unified theories.}
\end{quote}

Some of the GUTs are known to be quite complex; is this statement a premonition of the complexity of the landscape of string vacua and the multiverse?  

\section{Naturalness in Nonrelativistic Systems}

Somewhat paradoxically, gravity is perhaps the least-understood of all the fundamental interactions.  The possibility that gravity at short distances could behave nonrelativistically \cite{mqc,lif} has generated some interest, leading to a new perspective on the possible reconciliation of gravity with quantum mechanics, and on its short-distance completion in the quantum theory.  Since gravity plays a central role in the puzzles mentioned above, we are motivated to investigate technical naturalness for nonrelativistic systems.  As it turns out, this investigation reveals many surprises, even before we include gravity.  Let's start with a single scalar field in a flat nonrelativistic spacetime.  (After all, the interesting puzzles of gravity also involve scalars.)  Surely, quantum field theories of a single scalar field must have been exhaustively understood in the literature a long time ago?  Not so, as we will now see!

\subsection{Towards the Classification of Nonrelativistic Nambu-Goldstone Modes}

To focus our search for novel phenomena of naturalness, we choose the time-honored problem which universally emerges in the context of spontaneous symmetry breaking:  The problem of classifying the types of Nambu-Goldstone (NG) modes implied by the breaking of global continuous internal symmetries.    

In the relativistic case, the number and behavior of such NG modes is controlled by the simple, elegant and precise statement of Goldstone's theorem: One independent massless NG mode exists for each broken symmetry generator.  In nonrelativistic systems, with or without Galilean invariance, a similarly complete classification is not yet available.  Goldstone's theorem still exists, but it only says that there are NG modes and that they are gapless.  

A sensible strategy for this problem was proposed in \cite{wm1,wm2,hahi}: The classification will be accomplished if we can classify all effective field theories available to describe the NG mode dynamics.  Let us illustrate this method by focusing on nonrelativistic field theories in an ``Aristotelian spacetime'' $M$, of $D+1$ spacetime dimensions.  $M$ is topologically $\R^{D+1}$, and it is equipped with the flat metric -- but compared to either Lorentzian or Galilean spacetimes, the symmetries of $M$ are reduced to the ``Aristotelian symmetries'': The $SO(D)$ of spatial rotations, and the $\R^{D+1}$ of space and time translations.%
\footnote{Such spacetimes have often been called ``Lifshitz spacetimes'' in the hep-th literauture of recent years, but perhaps the term ``Aristotelian spacetime'' (as first introduced for spacetimes of this symmetry by Penrose \cite{penb} in 1968) is more appropriate.  We adopt this more accurate terminology from now on.}
The effective action, in terms of scalars $\pi^I(t,\bx)$, $I=1,\ldots K$, will be
\be
\label{efa}
\int dt\,d^D\bx\left(\Omega_J(\pi)\dot\pi^J+g_{IJ}(\pi)\dot\pi^I\dot\pi^J-
h_{IJ}(\pi)\p_i\pi^I\p_i\pi^J+\ldots\right).
\ee
The $\Omega$ term leads to a natural pairing of some of the $\pi^I$'s into canonical pairs; those will be called the ``Type B'' modes.  The other $\pi^I$'s do not participate in the $\Omega$ term, and they behave akin to the relativistic modes.  If this is the generic low-energy action, the classification of NG modes is simple: 

\begin{itemize}
\item Type A modes, characterized by dispersion relation $\omega\sim k$, with each such mode associated with a single broken symmetry generator.  

\item Type B modes, with dispersion $\omega\sim k^2$.  One such mode exists per each {\it pair\/} of broken symmetry generators paired by $\Omega_I$.  
\end{itemize}

Based on our intuition developed in relativistic systems, one may be tempted to conjecture that (\ref{efa}) is generic, and therefore everything else is fine-tuning.  This intuition is incorrect.  One can find explicit examples of interacting theories, in which the dispersion relations start at higher orders than quadratic, and the vanishing (or the smallness) of the lower-order terms is not spoiled by quantum corrections.  Consider for example the super-renormalizable linear sigma model of the scalar $\Phi$ in the $N$ representation of $O(N)$, in $2+1$ dimensions, defined by turning on a relevant self-interaction at the $z=2$ fixed point,%
\footnote{Other examples, leading to the same conclusions, can be found in \cite{msb}.}
with action
\be
\label{exab}
S_{\rm LSM}=\frac{1}{2}\int dt\,d^2\bx\left(\dot\Phi\cdot\dot\Phi-\p_i\p_j\Phi\cdot\p_i\p_j\Phi-c^2\p_i\Phi\cdot\p_i\Phi-m^4\Phi\cdot\Phi-\lambda(\Phi\cdot\Phi)^2\right).
\ee
Set $c^2=0$ at tree level.  The leading quantum contribution to $c^2$ comes from two loops, and was evaluated in \cite{msb} to be $\delta c^2=A(\lambda^2/m^4)$, where $A$ is a constant smaller than one, $\approx 0.125$.  This is quite small; how small?  Assume that there is a hidden symmetry, which is -- similarly to the constant shift symmetry in (\ref{epssc}) -- broken by some small amout $\varepsilon$ at some high momentum scale $\mu$, and protects the smallness of $c^2$. Then we expect, in complete analogy with (\ref{epssc}),
\be
\label{hidden}
m^4\sim\varepsilon\mu^4, \qquad\lambda\sim\varepsilon\mu^3, \qquad c^2\sim\varepsilon\mu^2. 
\ee
We can again obtain $\mu$ in terms of the physical couplings, $\mu\sim m/\lambda$, which predicts the small value of $c^2\sim\lambda^2/m^4$ if (\ref{hidden}) is to hold -- but this is precisely the behavior we found by the explicit loop calculation reported above!  Hence, if technical naturalness is upheld by this system, there must be a hidden symmetry behind (\ref{hidden}).  

\subsection{Polynomial Shift Symmetries}

This hidden symmetry, first identified in \cite{msb}, is the linear shift symmetry,
\be
\label{linshft}
\delta\phi(t,\bx)=a+a_ix^i.
\ee
The $c^2$ term is invariant under this symmetry, up to a total derivative, and therefore protected from (large) quantum corrections.%
\footnote{This symmetry (\ref{linshft}) is reminiscent of Galileons, but there are clear differences:  Galileons \cite{ggnrt} are relativistic theories, with scalars invariant under shifts linear in {\it spacetime\/} coordinates.  Since shifts linear in time are a part of the symmetry of the Galileons, their Hamiltonians and hence their dynamics are quite different from our nonrelativistic theories.  Moreover, in our nonrelativistic systems, the linear shift symmetries can be naturally extended to higher polynomial shifts; similar extensions in the relativistic context of the Galileons would lead to apparent violations of perturbative unitarity.}

(\ref{linshft}) can be easily generalized to higher polynomial shift symmetries%
\footnote{When the degree $P$ is greater than one, we can in principle split $a_{i_1\ldots i_P}$ into irreducible representations of $SO(D)$, and consider strictly smaller subgroups of polynomial shifts of degree $P$.  We will not deal with this refinement of the symmetry structure here.}
\be
\label{deltap}
\delta_P\phi(t,\bx)=a+a_ix^i+\ldots+a_{i_1\ldots i_P}x^{i_1}\ldots x^{i_P}.
\ee
This hierarchy of polynomial shift symmetries of degree $P$ leads to a refimenent of the classification of NG modes.  We obtain, in a technically natural way, two {\it towers of multicritical NG modes\/} \cite{msb}: 
\begin{itemize}
\item The Type A${}_n$ sequence, with dispersion $\omega\sim k^n$, with one mode per each broken symmetry generator.  
\item The Type B${}_{2n}$ sequence, with dispersion $\omega\sim k^{2n}$, with one mode per a {\it pair\/} of broken symmetry generators.  
\end{itemize}
Each of the two towers is labeled by an integer $n=1,2,\ldots$.  In addition, in the Type A case, we run into the multicritical generalization of the Coleman-Hohenberg-Mermin-Wagner (CHMW) theorem: There is a ``lower critical dimension'' of space at which the scalar becomes dimensionless, and its two-point function sensitive to an infrared regulator -- the NG mode with $n=D$ doesn't quite exist as a quantum object.  This would appear to limit the sensible unversality classes that can represent NG modes of spontaneous symmetry breaking to $n<D$; see, however, the next paragraph for a way out from this limitation.  

In the Type B case, there is no analog of the CHMW theorem, the dimension of the scalar field is always non-negative, and all the Type B${}_{2n}$ modes are free to serve as NG modes associated with spontaneous symmetry breaking.

\subsection{Naturalness of Cascading Hierarchies}

So far we focused on the classification of NG modes characterized by one Gaussian fixed point, protected by a polynomial shift symmetry of some degree $P$.  Since the polynomial shifts of degree $P'<P$ are contained in the shifts of degree $P$, the symmetries can be broken step by step, in a hierarchical fashion -- first, at some high scale $\mu$, from degree $P$ to degree $P'$; then at some lower scale $\mu'$ from $P'$ to $P''$; and so on, possibly all the way to the breaking of the constant shift symmetry.  This hierarchy $\mu\ll\mu'\ll\mu''\ldots$ produces a hierarchical sequence of crossover scales, which separate consecutive regimes with the Type A${}_n$, A${}_{n'}$, A${}_{n''}$, $\ldots$ behavior, with $n>n'>n''>\ldots$ (and similarly for Type B). Thus, the system exhibits a pattern of ``hierarchical symmetry breaking'', cascading naturally from higher to lower values of $n$.  Importantly, the technical naturalness of this mechanism will be protected even in the interacting theories, by the pattern of partial breakings of the polynomial shift symmetries.  

Note that this mechanism allows us to avoid the naive consequences of the multicritical version of the CHMW theorem:  The $n=D$ NG mode does exhibit a logarithmic propagator, and is sensitive to an IR cutoff -- but unlike in the relativistic case, the system can have a natural physical IR cutoff, represented by any of the scales of crossover from $n=D$ to $n<D$, without losing the constant shift symmetry required of the NG modes.  In this sense, spontaneous symmetry breaking associated with such naturally regulated $n=D$ NG modes is possible.  

\section{Interacting Theories with Polynomial Shift Symmetries}

Any generic interaction will generally break all polynomial shift symmetries all the way to the constant shift guaranteed by Goldstone's theorem.  This is the case in the example (\ref{exab}), and those of \cite{msb}.  Now that we know that the new kind of symmetry is possible and can be realized by free-field fixed points, we can turn the argument around, and ask:  Given a polynomial shift symmetry of degree $P$ in $D+1$ spacetime dimensions, what is the most general Lagrangian which respects this symmetry?  It is natural to organize this question by the growing dimensions of the scalar composite operators that can appear in the Lagrangian.  The first question is:  given $P$, $D$, and the number of fields $n$, what is the lowest-dimension operator that can appear in the Lagrangian, and is invariant up to a total derivative?

\subsection{Invariants of Polynomial Shift Symmetries}

The questions just posed are essentially cohomological in nature.  The classification of possible terms $\CL$ in the Lagrangian, made of $n$ copies of $\phi$, containing $2\Delta$ spatial derivatives, satisfying the invariance relation
\be
\label{cohom}
\delta_P\CL=\p_i\CL_i
\ee
(with $\delta_P$ acting as in (\ref{deltap})), and not equivalent by integration by parts, defines a sequence of cohomology groups, labeled by the integers $P$, $D$, $n$, and $\Delta$.  The evaluation of these cohomology groups is an intriguing mathematical problem; some partial answers can be found in \cite{pol}.

\subsection{Polynomial Shift Invariants and Graph Theory}

The operations defining this sequence of cohomology groups has an intriguing representation in the language of graph theory.  Using this representation, one thus defines a new infinite sequence of cohomology groups associated with operations on graphs.  There is a larger set of graph-theory tools needed to set up (\ref{cohom}) and to classify its solutions (see \cite{pol}); fortunately, only a pleasingly simple toolset is needed to state the results.  Each monomial in $\CL$ is made of products of $\p_i\ldots\p_j\phi$'s, and can be depicted as follows:  Represent each copy of $\phi$ by a dot (i.e., a vertex of the graph), and each pair $\p_i\ldots\p_i\ldots$ of contracted derivatives by a line.  The line is attached to the two vertices representing the $\phi$'s that the $\p_i$'s are acting on.  This language is surprisingly efficient not only for deriving theorems about our invariants, but also for stating the results.  

\subsection{Examples}

We begin with invariants with linear shift symmetries: At $n$ points and in dimensions $D>n-1$, the lowest-dimension invariant occurs at $\Delta=n-1$, and it is unique.  This is proven in \cite{pol}, for any $n$ and without any additional simplifying assumptions common in the analogous statements in the Galileon literature.  Moreover, the invariant has an intriguing and surprisingly simple representation in the graph theory language:  It is simply given by the sum over all trees, with equal weight, with the vertices treated as labeled or distinguishable.  For example, the lowest-dimension 4-point $P=1$ invariant has six derivatives, and is given by 
\be
{\hbox{\includegraphics[angle=0,width=2.5in]{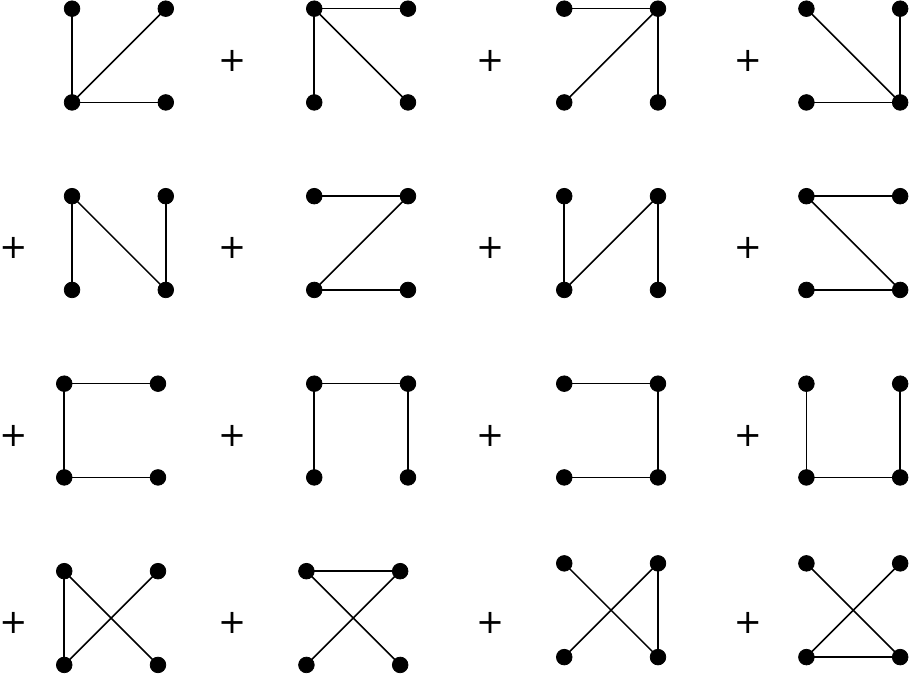}}}\ \ .
\ee
When we treat the vertices as indistinguishable, this sum becomes
\be
4\ \ \vcenter{\hbox{\includegraphics[angle=0,width=.3in]{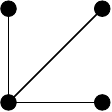}}}+12\ \vcenter{\hbox{\includegraphics[angle=0,width=.3in]{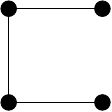}}}\ =4\p_i\p_j\p_k\phi\,\p_i\phi\,\p_j\phi\,\p_k\phi+12\p_i\phi\,\p_i\p_j\phi\,\p_j\p_k\phi\,\p_k\phi.
\ee
It is at present unknown what underlies this surprising simplicity of the lowest-dimension $n$-point 1-invariants in the language of graph theory.  

The graph theory techniques become indispensable for higher values of $P$ and $\Delta$.  To illustrate, here is the unique, lowest-derivative 4-point, $P=5$ invariant, with vertices treated as indistinguishable:

\be
{\hbox{\includegraphics[angle=0,width=3.8in]{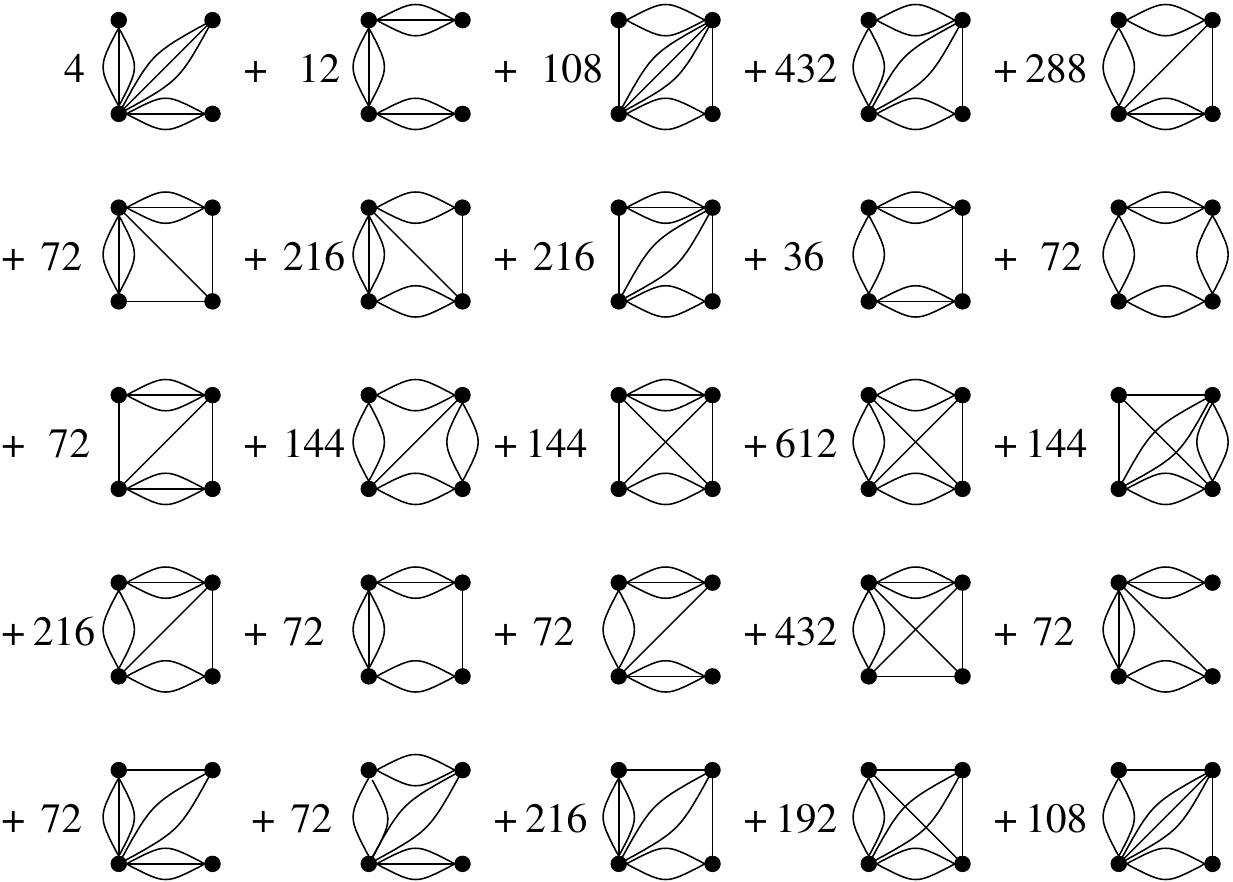}}}\ \ .
\ee
Each of the $16^3=4,096$ terms depicted in this invariant has eighteen derivatives distributed aroud the four $\phi$'s.  In the graph language, this invariant is formed via a graph-theory superposition of three copies of the $P=1$ $\Delta=3$ invariant; see \cite{pol} for details and proofs.

\section{Back to Physics}

Having seen a glimpse of the surprisingly rich mathematical structure of polynomial shift symmetries and their invariants, we now return back to physics, and look for applications involving our technically natural NG modes.  

In \S 1 we mentioned the puzzle of linear resistivity in the strange metal phase above $T_c$ in high-$T_c$ superconductors.  Can we find a mechanism which causes resistivity in metals to scale linearly with temperature, in a technically natural way?  

In the standard BCS theory, superconductivity is caused by the electron-phonon interactions.  In the metallic regime above $T_c$, the resistivity caused by these interactions is given by the Bloch-Gr\"uneisen formula, leading to the well-known $T^5$ scaling at sufficiently low $T$.  In contrast, the temperature dependence of resistivity $\rho(T)$ in many high-$T_c$ superconductors looks qualitatively like this:
\[
\includegraphics[angle=0,width=1.5in]{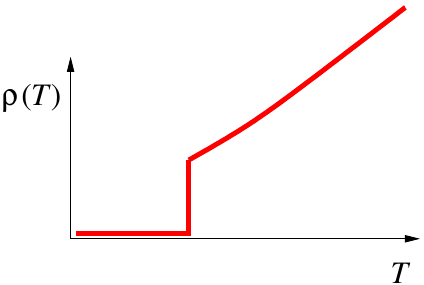}
\]
This behavior persists over a surprisingly large range of scales, and it is also quite robust under various deformations of the system, such as doping.  How to explain this, in a technically natural way?  And what is the ``superglue'' that serves to create pairing in these systems?  The electron-phonon interactions yield $\rho(T)\sim T^5$, and appear to be out.  Electron-electron interactions give $T^2$, and the interactions with impurities lead to $\rho(T)\sim {\rm const}$.  Again, none of them give resistivity linear in temperature.%
\footnote{$\rho(T)\sim T$ does emerge naturally and simply from the electron-ion interactions in the {\it high-temperature\/} regime, with $T$ above the Debye temperature -- but here we are interested in a much lower range of $T$, well below the Debye temperature.}  
Perhaps some very novel, exotic mechanism is taking place?  
\footnote{Among the most popular such ideas is quantum criticality, which can robustly lead to linear resistivity given the absence of scales near the purported critical point.}

This is a situation not dissimilar to the puzzles of naturalness involving gravity, which are also often interpreted as a sign that revolutionary new ideas about quantum gravity are needed.  However, it is interesting to go against this flow, and revisit the traditional mechanisms with the added understanding of nonrelativistic systems.  For gravity, this strategy was initiated in \cite{mqc,lif}.  Following this path in the case of superconductivity leads us to revisit the time-tested BCS theory, and to return to the electron-phonon interactions as the cause of electron pairing.  

Needless to say, acoustic phonons represent a truly classic example of a NG mode, associated with breaking of translational invariance.  Having seen that NG modes with higher-order dispersion relations are technically natural, we now ask what happens in the BCS theory when the role of the phonons is played by the Type A${}_n$ NG modes of \cite{msb,cmu}, with $n>1$.

\subsection{Multicritical Phonons and the Debye Model}

We will address this question in the simplest, ``jellium'' model of a metal in the Aristotelian spacetime of $3+1$ dimensions, replacing the ionic lattice by a uniform medium, taking the Fermi surface of the electrons to be spherically symmetric.  In this model, all umklapp processes are neglected; only the longitudinal component of the phonons remains coupled, and behaves as a true scalar.  With this scalar being of Type A${}_1$, this would be the simplest model of a conventional BCS superconductor.  We now make the crucial assumption that the phonons are NG modes of Type A${}_n$, with $n>1$.%
\footnote{The scalar collective mode that we refer to as the ``phonon'' does not have to come literally from the lattice vibrations -- it could be, for example, magnetic in origin.  We refer to this mode as the ``phonon'' for simplicity, regardless of its microscopic origin.}  

In the Gaussian approximation, our phonons will be described by a ``multicritical Debye model'' \cite{res}.  Using the hydrodynamic description, we write the fluid density as $\rho(t,\bx)=\rho_0+Q(t,\bx)$, where $Q$ is the deviation from a uniform isotropic background $\rho_0$.  We assume the velocity $\bv(t,\bx)$ of the fluid to describe a gradient flow, 
\be
v^i(t,\bx)=\p_i f(t,\bx).
\ee
This ensures that the transverse phonons are absent; the purely longitudinal phonons are then described by the canonical pair $f$ and $Q$, with $[f(t,\bx),Q(t,\bx')]=-i\delta^3(\bx-\bx')$.  In the Gaussian approximation, the Hamiltonian is
\be
H_{\rm ph}=\int d^3\bx\left(\frac{1}{2}\rho_0\bv^2+\CV(Q)\right).
\ee
In order for this system to describe the multicritical Type A${}_n$ phonon, we assume that $\CV$ is dominated by 
\be
\CV=\frac{1}{2}\zeta_n^2\,\p_{i_1}\ldots\p_{i_{n-1}}Q\,
\p_{i_1}\ldots\p_{i_{n-1}}Q.
\ee
This yields phonon excitations with the Type A${}_n$ dispersion relation $\omega^2=\zeta_n^2 k^{2n}$.%
\footnote{Terms with lower integer values of $n$ are also allowed, as long as they are sufficiently small so that they only become important at momenta much lower than those of interest.} 

In the standard Debye model, there is a natural short-distance cutoff:  The momenta cannot exceed $k_D$, and the frequencies are correspondingly cut off at the Debye frequency $\omega_D$.  In the multicritical Debye model with Type A${}_n$ phonons, similarly, the momenta cannot exceed $k_D$, and $\omega$ goes only up to the modified Debye frequency $\tilde\omega_D$, whose value depends on the coefficients of the phonon dispersion relation.  In both cases, the value of $\omega_D$ or $\tilde\omega_D$ is determined by requiring that the total number of modes in the system is equal to that of the underlying lattice.  An interesting special case is $n=D$, the case of the lower critical dimension (cf.\ Fig.~1).  

\begin{figure}[h]
\centering
\includegraphics[angle=0,width=4.5in]{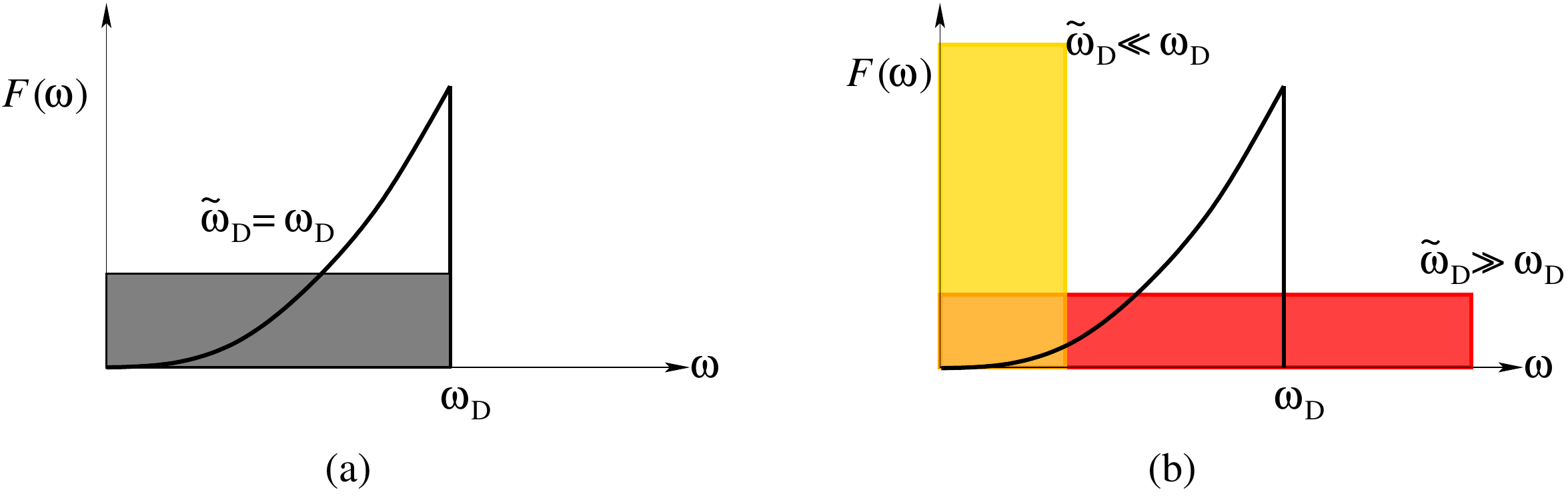}
\caption{The density of states $\varrho(\omega)$ as a function of frequency 
in the multicritical Debye model with $n=3$ in $3+1$ dimensions (the lower critical case), compared to the density of states in the standard Debye model. {\bf (a):} 
the case of $\tilde\omega_{\rm D}=\omega_{\rm D}$; {\bf (b):} the regimes with 
$\tilde\omega_{\rm D}\ll\omega_{\rm D}$ and 
$\tilde\omega_{\rm D}\gg\omega_{\rm D}$, assuming fixed $k_{\rm D}$.}
\end{figure}

In the conventional systems with phonons, the terms of lowest order in $k$ dominate the low-energy dispersion relation, with the size of all higher-order terms generically suppressed hierarchically.  How can we turn this into a system with multicritical phonons?  There are two interesting options how a sizable hierarchy can be opened:  Either by suppressing the lower-order terms, or by enhancing the higher-order term.  
Phenomenologically, the second option appears interesting for superconductivity, suggesting the following behavior of the density of states:  
\[
\includegraphics[angle=0,width=2in]{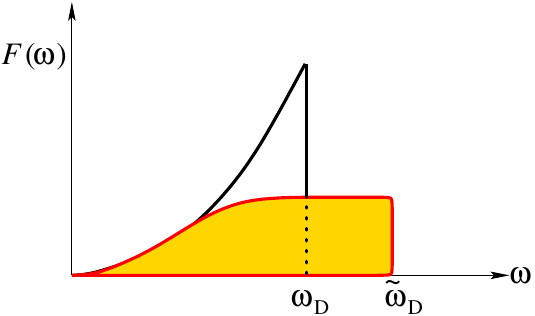}
\]
We will adopt this as our model of the acoustic phonons in a metal.  Such phonons are an example of the cascading hierarchy of scales discussed in \S 2.3 (and in \cite{cmu}):  The $n=3$ dispersion dominates in the regime below $\tilde\omega_D$, and changes to $n=1$ at a hierarchically much smaller crossover scale $\omega_\otimes\ll\tilde\omega_D$ (possibly also with an intermediate crossover to $n=2$).

\subsection{Coupling to the Fermi Surface}

The electrons $\Psi(t,\bx)$ are assumed for simplicity to have the spherical Fermi surface.  We take the coupling of the phonons to the fermions to be
\be
g\int dt\,d^3\bx\,Q\Psi^\dagger\Psi\equiv g\int dt\,d^3\bx\,\p_iQ_i\Psi^\dagger\Psi.
\ee
This is the standard coupling between the electron Fermi liquid and the phonon excitations of the lattice, given in terms of the displacement field $Q_i(t,\bx)$, with our $Q=\p_iQ_i$.  This standard coupling breaks the polynomial shift symmetry of the phonon system all the way to the constant shifts of $Q_i$, but does so in a way which does not violate the hierarchy of scales $\omega_\otimes\ll\tilde\omega_D$.  

\subsection{Resistivity in Strange Metals}

We use Bloch-Boltzmann kinetic theory of transport to calculate the resistivity due to the interaction between the electrons and our multicritical phonons.  The result generalizes the Bloch-Gr\"uneisen formula, which gives resistivity $\rho$ in terms of the {\it relaxation time\/} $\tau_{\rm tr}(\varepsilon_F)$, evaluated at the Fermi surface (see, e.g., \cite{pottier}, Ch.~8), 
\be
\rho\sim\frac{1}{\tau_{\rm tr}(\varepsilon_F)}\sim\int_\muir^{k_D}|g_k|^2n(k)k^2\,k\,{\rm d}k.
\ee
Here $n(k)=1/[\exp(\omega_k/T)-1]$ is the phonon distribution function, $g_k\sim g\,k/\sqrt{\omega_k}$ the electron-phonon vertex, and $\muir$ is an infrared regulator, which can be set to zero in the absence of infrared divergences.%
\footnote{When the phonons are at the lower critical dimension, there are IR divergences that are automatically regulated with $\omega_\otimes>0$ serving as the natural IR regulator, much as in \S 2.3.}
In the range of temperatures of interest, well below the Debye temperature $\Theta\sim k_D^n$, this formula gives resistivity which depends on $T$ and $n$ as
\be
\label{scalingthree}
\rho(T)\sim T^{(6-n)/n}.
\ee
In this standard case of $n=1$, this reproduces the standard Bloch-Gr\"uneisen scaling $\rho\sim T^5$.  More interestingly, in the case of the multicritical phonons at their lower critical dimension, $n=3$, we obtain $\rho\sim T$ (accompanied by $T\log T$ corrections) -- the electron-phonon interaction is capable of explaining resistivity linear in temperature!

The calculation can be repeated in any number of spatial dimensions $D$, generalizing (\ref{scalingthree}) to 
\be
\label{scalingdee}
\rho(T)\sim T^{(3+D-n)/n}.
\ee
The interesting cases are those with $D=2$ and 3: Type A{}$_n$ multicritical phonons with $n\leq D$ suggest resistivity going as $T^5$, $T^2$ (which matches thescaling of the electron-electron contribution) or $T$ in three spatial dimensions, and $T^4$ or $T^{3/2}$ in two dimensions.  

\section{Conclusions and Future Prospects}

Investigating the role of technical naturalness in nonrelativistic systems with Aristotelian spacetime symmetries has led to many surprises, even in the simplest cases of single scalar fields, often challenging our naive intuition based on relativistic systems.  These surprises should be finding finding their first applications in those fields of physics which naturally respect the Aristotelian symmetries:  The most obvious is condensed matter, where we have found a technically natural explanation for resistivity linear in temperature in metals with multicritical phonons.  Another field which shares similar spacetime symmetries with a preferred rest frame is cosmological inflation; applications in this area are therefore also likely.  

Eventually, we are hoping that the lessons learned in the nonrelativistic arena will give new insights to the ``fundamental'' puzzles of naturalness in high-energy physics and gravity:  The cosmological constant problem and the Higgs mass hierarchy problem.  If this many surprises and such unexpectedly rich mathematical and physical structure appear already in the case of a single scalar, one can only imagine how many surprises are still hidden in nonrelativistic systems with gauge symmetries, in particular in nonrelativistic gravity.  

\acknowledgments

It is a pleasure to thank Tom Griffin, Kevin Grosvenor, Ziqi Yan and Chris Mogni for the many fruitful and enjoyable collaborations on the results summarized in these notes.  This paper is based on talks and lectures given at the 2nd LeCosPA Symposium on {\it Everything About Gravity} at National Taiwan University, Taipei, Taiwan (December 2015); at the {\it International Conference on Gravitation and Cosmology}, KITPC, Chinese Academy of Sciences, Beijing, China (May 2015); at the Symposium {\it Celebrating 100 Years of General Relativity}, Guanajuato, Mexico (November 2015); and at the {\it 54.\ Internationale Universit\"atswochen f\"ur Theoretische Physik}, Schladming, Austria (February 2016).  I wish to thank Pisin Chen; Rong-Gen Cai and Remo Ruffini; Gustavo Niz; and Daniel Litim and Willibald Plessas, for their wonderful organization and hospitality in Taipei, Beijing, Guanajuato and Schladming, respectively.  This work has been supported by NSF Grant PHY-1521446 and the Berkeley Center for Theoretical Physics.

\bibliographystyle{JHEP}
\bibliography{nrr}
\end{document}